\documentclass[20pt]{article}
\usepackage{indentfirst}
\usepackage{amsmath}
\usepackage{amssymb}
\usepackage{latexsym}
\usepackage{amsmath}
\usepackage{lineno}
\usepackage{graphicx}
\usepackage{float}
\usepackage{subfigure}
\usepackage{geometry}
\usepackage{graphics}
\usepackage{subfigure}
\usepackage{graphicx}
\usepackage{multirow}
\usepackage{booktabs}
\usepackage{hyperref}
\usepackage{cite}
\makeatletter

\newcommand{\be}{\begin{equation}}
\newcommand{\ee}{\end{equation}}

\begin{document}
\begin{center}
\large {\bf Dirac equation of spin particles and tunneling radiation from an Kinnersly black hole}
\end{center}

\begin{center}
Guo-Ping Li $^1$
 $\footnote{E-mail: Corresponding author, gpliphys@yeah.net}$
Zhong-Wen Feng $^2$
 $\footnote{E-mail:  \texttt{zwfengphy@163.com}}$
Hui-Ling Li $ ^{1, 3}$
 $\footnote{E-mail: \texttt{LHL51759@126.com}}$
Xiao-Tao Zu ${^1}$
 $\footnote{E-mail:  \texttt{xtzu@uestc.edu.cn}}$
\end{center}

\begin{center}
\textit{1. School of Physical Electronics, University of Electronic Science and Technology of China, Chengdu, 610054, China\\
2. Department of Astronomy, China West Normal University, Nanchong, 637009, China\\
3. College of Physics Science and Technology, Shenyang Normal University, Shenyang, 110034, China}
\end{center}

\noindent
{\bf Abstract:} In curved space-time, Hamilton-Jacobi equation is a semi-classical particle's motion equation, which plays an important role in the research of black hole physics. In this paper, starting from Dirac equation describing the spin 1/2 fermion and Rarita-Schwinger equation describing the spin 3/2 fermion respectively, we derive a Hamilton-Jacobi equation of the non-stationary spherically symmetric gravitational field background. Furthermore, the quantum tunneling behavior of a charged spherically  symmetric black hole is investigated by using this Hamilton-Jacobi equation. The result shows that the Hamilton-Jacobi equation is helpful for people to understand the thermodynamic properties and the radiation characteristics of a black hole.


\section{Introduction}
 \label{section1}
\setlength{\parindent}{2em}
In 1974, considering quantum effects, Hawking proved that a black hole has the thermal radiation \cite{1}. After that, people have carried out a series of research for various types of black hole radiation \cite{2,3,4,5,7}. About the origin of Hawking radiation, a kind of viewpoint is that the virtual particles inside a black hole via the quantum tunneling effect reach the event horizon of a black hole and materialize real particles, and radiate out, which called Hawking thermal radiation \cite{7+}. (About the source of Hawking radiation, a common viewpoint believes that, due to vacuum fluctuation near the event horizon where a virtual particle pair could be created or annihilated, a negative energy particle falls in since there exists a negative energy orbit inside black hole, however the positive energy particle left outside is radiated to infinite, which causes Hawking radiation \cite{8}). In 2000, Parikh and Wilczek \emph{et al}. put forward that a quantum tunneling theory to study the thermal radiation of a black hole. By taking into account the background change before and after radiation particles tunneling, they have carried out modification to the previous tunneling probability \cite{9}. Thus, in recent years, a series of significant studies have been made on the tunneling radiation of black holes \cite{10}. Zhang and Zhao \emph{et al.} have developed this tunneling theory, and studied the relationship between the tunneling radiation and the black hole entropy, which provides a reasonable explanation for information loss paradox of a black hole. Further research of Lin and Yang \emph{et al.} showed that the tunneling rate of the event horizon from a dynamic black hole is not only related to the black hole entropy, but also related to an integral function \cite{11}. Therefore, information loss paradox of a black hole still needs further research. According to the literature \cite{12,13,131+,131a+,131b+,13+,14+,15+,16+,17+,add1}, the Hamilton-Jacobi equation in curved space-time is the basic equation describing the dynamic characteristics of all kinds of particles. For the Dirac equation describing the spin particle and the equation describing the spin 3/2 particle, the Dirac equation and the tunneling radiation of a stationary black hole have been studied. However, a black hole in the universe is due to radiation, accretion, merging and other reasons, which leads to be dynamic change of the black hole. Therefore, it is a practical significance to study the tunneling radiation characteristics of a dynamic black hole.

For a dynamic black hole, the Dirac equation describing the spin $1/2$ and the $3/2$ particles is more complex than that of a stationary black hole. In this paper, according to the space-time linear element of a Kinnersley black hole, using the advanced Eddington coordinate to represent its dynamic characteristics, we study the tunneling radiation of spin $1/2$ and $3/2$ particles in this space-time background.

The rest of the paper is organized as follows. Due to the Dirac equation for spin $1/2$ particle and the Rarita-Schwinger equation for spin $3/2$ particle, we derived the Hamilton-Jacobi equation in Section~\ref{section2}. In Section~\ref{section3}, according to the Hamilton-Jacobi equation, the fermion tunnelling behavior in a non-stationary Kinnersley black hole is addressed. Section~\ref{section4} is devoted to our discussion and conclusion.

\section{Dirac equation of spin particles and the Hamilton-Jacobi equation}
 \label{section2}
\setlength{\parindent}{2em} The Dirac equation for spin $1/2$ particle can be described by \cite{14}
\begin{eqnarray}
\gamma ^\mu  D_\mu  \Psi  + \left( {{m \mathord{\left/
 {\vphantom {m \hbar }} \right.
 \kern-\nulldelimiterspace} \hbar }} \right)\Psi  = 0,
 \label{eq1}
\end{eqnarray}
where
\begin{eqnarray}
D_\mu   = \partial _\mu   + \frac{1}{2}i\Gamma _\mu ^{\alpha \beta } \Pi _{\alpha \beta } ,
 \label{eq2}
\end{eqnarray}
\begin{eqnarray}
\Pi _{\alpha \beta }  = \frac{i}{4}\left[ {\gamma ^\alpha  ,\gamma ^\beta  } \right],
 \label{eq3}
\end{eqnarray}
here the relationship between gamma matrices and the space-time metric is
\begin{eqnarray}
\left\{ {\gamma ^\mu  ,\gamma ^\nu  } \right\} = 2g^{\mu \nu } I,
 \label{eq4}
\end{eqnarray}
Before solving the equation Eq.~(\ref{eq1}), one must know the space-time background. Let us consider a dynamic Kinnersley black hole of arbitrarily variably accelerated rectilinear motion. In the advanced Eddington coordinate, the space-time linear element of the black hole can be written as \cite{15}
\begin{eqnarray}
 ds^2  = & - & \left( {1 - 2ar\cos \theta  - r^2 f^2  - 2Mr^{ - 1} } \right)d\upsilon ^2  - 2d\upsilon dr + 2r^2 fd\upsilon d\theta
\nonumber \\
  & - &r^2 d\theta ^2 - r^2 \sin ^2 \theta d\varphi ^2 ,
 \label{eq5}
\end{eqnarray}
where  $f =  - a\left( \upsilon  \right)\sin \theta$, $M = M\left( \upsilon  \right)$. $a\left( \upsilon  \right)$ stands for the magnitude of acceleration. $M\left( \upsilon  \right)$ is the mass of the black hole.  From Eq.~(\ref{eq2}), one gets the metric determinant and the non-zero components of the inverse metric tensor
\begin{eqnarray}
g =  - r^4 \sin ^2 \theta,
 \label{eq6}
\end{eqnarray}
and
\begin{eqnarray}
g^{01}  = g^{10}  =  - 1,g^{11}  =  - \left( {1 - 2ar\cos \theta  - 2Mr^{ - 1} } \right),g^{12}  = g^{21}  = f,g^{22}  =  - {1 \mathord{\left/
 {\vphantom {1 {r^2 }}} \right.
 \kern-\nulldelimiterspace} {r^2 }}, g^{33}=r^{-2}\sin ^{-2} \theta.
 \label{eq7}
\end{eqnarray}
According to Eq.~(\ref{eq5}) - Eq.~(\ref{eq7}), the gamma matrices of Eq.~(\ref{eq4}) can be constructed as follows
\begin{eqnarray}
\gamma ^\upsilon   =\sqrt { - \left( {1 - 2ar\cos \theta  - 2Mr^{ - 1} } \right)^{ - 1} } \left[ {i\left( {\begin{array}{*{20}c}
   I & 0  \\
   0 & { - I}  \\
\end{array}} \right) + \left( {\begin{array}{*{20}c}
   0 & {\sigma ^3 }  \\
   {\sigma ^3 } & 0  \\
\end{array}} \right)} \right],
 \label{eq8}
\end{eqnarray}
\begin{eqnarray}
\gamma ^r  =\sqrt { - \left( {1 - 2ar\cos \theta  - 2Mr^{ - 1} } \right)} \left( {\begin{array}{*{20}c}
   0 & {\sigma ^3 }  \\
   {\sigma ^3 } & 0  \\
\end{array}} \right),
 \label{eq9}
\end{eqnarray}
\begin{eqnarray}
\gamma ^\theta   = \frac{f}{{\sqrt { - \left( {1 - 2ar\cos \theta  - 2Mr^{ - 1} } \right)^{ - 1} } }}\left( {\begin{array}{*{20}c}
   I & 0  \\
   0 & { - I}  \\
\end{array}} \right) + \sqrt {\frac{{\chi \left( { - {1 \mathord{\left/
 {\vphantom {1 {r^2 }}} \right.
 \kern-\nulldelimiterspace} {r^2 }}} \right) - f^2 }}{{ - \left( {1 - 2ar\cos \theta  - 2Mr^{ - 1} } \right)^{ - 1} }}} {\rm{ }}\left( {\begin{array}{*{20}c}
   0 & {\sigma ^1 }  \\
   {\sigma ^1 } & 0  \\
\end{array}} \right),
 \label{eq10}
\end{eqnarray}
\begin{eqnarray}
\gamma ^\varphi   = \sqrt {r^{ - 2} \sin ^{ - 2} \theta } \left( {\begin{array}{*{20}c}
   0 & {\sigma ^2 }  \\
   {\sigma ^2 } & 0  \\
\end{array}} \right),
\label{eq11}
\end{eqnarray}
and in gamma matrices Eq.~(\ref{eq8}) - Eq.~(\ref{eq11}), the Pauli matrices are
\begin{eqnarray}
\sigma ^1  = \left( {\begin{array}{*{20}c}
   0 & 1  \\
   1 & 0  \\
\end{array}} \right),\sigma ^2  = \left( {\begin{array}{*{20}c}
   0 & { - i}  \\
   i & 0  \\
\end{array}} \right),\sigma ^3  = \left( {\begin{array}{*{20}c}
   1 & 0  \\
   0 & { - 1}  \\
\end{array}} \right),
\label{eq12}
\end{eqnarray}
where, $\chi = - \left( {1 - 2ar\cos \theta  - 2Mr^{ - 1} } \right)$. The wave function is expressed as
\begin{eqnarray}
\Psi  = \zeta \exp \left( {\frac{i}{\hbar }S} \right),
\label{eq13}
\end{eqnarray}
Where $S$ is the main function, the coefficient term can be decomposed into
\begin{eqnarray}
\zeta  = \left( {\begin{array}{*{20}c}
   A  \\
   B  \\
\end{array}} \right).
\label{eq14}
\end{eqnarray}
Substituting Eq.~(\ref{eq12}) - Eq.~(\ref{eq13}) into Eq.~(\ref{eq11}) and neglecting the terms with $\hbar$, one can get a matrix equation, i.e.
\begin{eqnarray}
\left( {\begin{array}{*{20}c}
   C & D  \\
   D & G  \\
\end{array}} \right)\left( {\begin{array}{*{20}c}
   A  \\
   B  \\
\end{array}} \right) = 0,
\label{eq15}
\end{eqnarray}
In Eq.~(\ref{eq15}), $C$, $D$ and $G$ are closely related to Eq.~(\ref{eq5})-Eq.~(\ref{eq12}). The expressions are respectively,
\begin{eqnarray}
C = m - \sqrt { - \left( {1 - 2ar\cos \theta  - 2Mr^{ - 1} } \right)^{ - 1} } \frac{{\partial S}}{{\partial \upsilon}}I - \frac{f}{{\sqrt { - \left( {1 - 2ar\cos \theta,  - 2Mr^{ - 1} } \right)^{ - 1} } }}\frac{{\partial S}}{{\partial \theta }}I,
\label{eq16}
\end{eqnarray}
\begin{eqnarray}
 D & = &i\left[ {\sqrt { - \left( {1 - 2ar\cos \theta  - 2M r^{ - 1} } \right)^{ - 1} } \frac{{\partial S}}{{\partial \upsilon }} + \sqrt { - \left( {1 - 2ar\cos \theta  - 2Mr^{ - 1} } \right)} \frac{{\partial S}}{{\partial r}}} \right]\sigma ^3
\nonumber \\
&  +& i\sqrt {\frac{{ - \left( {1 - 2ar\cos \theta  - 2Mr^{ - 1} } \right)^{ - 1} \left( { - {1 \mathord{\left/
 {\vphantom {1 {r^2 }}} \right.
 \kern-\nulldelimiterspace} {r^2 }}} \right) + f^2 }}{{ - \left( {1 - 2ar\cos \theta  - 2Mr^{ - 1} } \right)^{ - 1} }}} \frac{{\partial S}}{{\partial \theta }}\sigma ^1  + i\sqrt {r^{ - 2} \sin ^{ - 2} \theta } \frac{{\partial S}}{{\partial \phi }}\sigma ^2 ,
\label{eq17}
\end{eqnarray}
\begin{eqnarray}
G = m + \sqrt { - \left( {1 - 2ar\cos \theta  - 2Mr^{ - 1} } \right)^{ - 1} } \frac{{\partial S}}{{\partial \upsilon}}I + \frac{f}{{\sqrt { - \left( {1 - 2ar\cos \theta  - 2Mr^{ - 1} } \right)^{ - 1} } }}\frac{{\partial S}}{{\partial \theta }}I,
\label{eq18}
\end{eqnarray}
Because of $C D  = D C$, in order to obtain non-trivial solutions, it is necessary to have
\begin{eqnarray}
\det \left( {CG - DD} \right) = 0,
\label{eq19}
\end{eqnarray}
Taking into account the anti-commute relationship of Pauli matrices
\begin{eqnarray}
\left\{ {\sigma ^\mu  ,\sigma ^\nu  } \right\} = 2\delta _{\mu \nu },
\label{eq20}
\end{eqnarray}
According to Eq.~(\ref{eq16}) - Eq.~(\ref{eq18}), we can concretely express Eq.~(\ref{eq19}) as
\begin{eqnarray}
 - \left( {1 - 2ar\cos \theta  - 2Mr^{ - 1} } \right)\left( {\frac{{\partial S}}{{\partial r}}} \right)^2  - 2\frac{{\partial S}}{{\partial r}}\frac{{\partial S}}{{\partial \upsilon }} + 2f\frac{{\partial S}}{{\partial r}}\frac{{\partial S}}{{\partial \theta }} - \frac{1}{{r^2 }}\left( {\frac{{\partial S}}{{\partial \theta }}} \right)^2  + \frac{1}{{r^2 \sin ^2 \theta }}\left( {\frac{{\partial S}}{{\partial \phi }}} \right)^2  + m^2  = 0
\label{eq21}
\end{eqnarray}
This is the kinetic equation that only for  a spin $1/2$ particle. That is to say, we derive the equation Eq.~(\ref{eq21}) from the equation Eq.~(\ref{eq1}). Eq.~(\ref{eq21}) is just the Hamilton-Jacobi equation of the dynamic Kinnersley black hole of the variably accelerated rectilinear motion, which is described by the equation Eq.~(\ref{eq7}). One can simplify it as $g^{\mu \nu } \left( {{{\partial S} \mathord{\left/ {\vphantom {{\partial S} {\partial \mu }}} \right. \kern-\nulldelimiterspace} {\partial \mu }}} \right)\left( {{{\partial S} \mathord{\left/
 {\vphantom {{\partial S} {\partial \nu }}} \right. \kern-\nulldelimiterspace} {\partial \nu }}} \right) + m^2  = 0$.

At the same time, it is needed to explain that, the dynamical behavior of the spin $3/2$ particles is described by Rarita-Schwinger equation of the curved space-time. Its concrete expression is
\begin{eqnarray}
\gamma ^\mu  D_\mu  \Psi _\nu   + \left( {{m \mathord{\left/
 {\vphantom {m \hbar }} \right.
 \kern-\nulldelimiterspace} \hbar }} \right)\Psi _\nu   = 0,
\label{eq22}
\end{eqnarray}
\begin{eqnarray}
\gamma ^\mu  \Psi _\nu   = 0,
\label{eq23}
\end{eqnarray}
where every $\Psi _\nu$ is a matrix. The expressions Eq.~(\ref{eq2}) and Eq.~(\ref{eq23}) together can determine every wave function. In the semi-classical theory, the wave function in the equation Eq.~(\ref{eq22}) can be expressed as
\begin{eqnarray}
\Psi_\nu  = \left( {\begin{array}{*{20}c}
   {A_\nu  }  \\
   {B_\nu  }  \\
\end{array}} \right)\exp \left( {\frac{i}{\hbar }S} \right),
\label{eq24}
\end{eqnarray}
where $A = \left( {\begin{array}{*{20}c}   {a_\nu  } & {b_\nu  }  \\\end{array}} \right)^{TM}$ and $B = \left( {\begin{array}{*{20}c} {b_\nu  } & {d_\nu  }  \\
\end{array}}\right)^{TM}$ and $a_\upsilon$,$b_\upsilon$,$c_\upsilon$,$d_\upsilon$ express the corresponding matrix. Thus, for the spin $3/2$ particles, the gamma matrix can be chosen to be consistent with Eq.~(\ref{eq9}) - Eq.~(\ref{eq11}). In the same way, we can still discuss the tunneling radiation of spin $3/2$ particles in the expression~(\ref{eq7}), which describes a dynamic Kinnersley black hole of the variably accelerated rectilinear motion. In the semi classical approximation, we can still get
\begin{eqnarray}
\left( {\begin{array}{*{20}c}
   C & D  \\
   D & G  \\
\end{array}} \right)\left( {\begin{array}{*{20}c}
   {A_\nu  }  \\
   {B_\nu  }  \\
\end{array}} \right) = 0,
\label{eq25}
\end{eqnarray}
In the above expression $C$, $D$ and $G$ are shown such as expressions Eq.~(\ref{eq16}) - Eq.~(\ref{eq18}). Taking into account non-trivial solution conditions in Eq.~(\ref{eq25}), we can obtained
\begin{eqnarray}
\det \left( {CG - DD} \right) = 0,
\label{eq26}
\end{eqnarray}
By the calculation, the Hamilton-Jacobi equation which is consistent with the Eq.~(\ref{eq21}) can still be obtained. So, for the Rarita-Schwinger equation of spin $3/2$ fermion particles, we can also obtain the dynamic equation Eq.~(\ref{eq21}). This further shows that the Hamilton-Jacobi equation of curved space-time is a basic dynamic equation of the radiation particles.

\section{Hamilton-Jacobi equation and the tunneling from Non-Stationary Kinnersly black hole}
 \label{section3}
\setlength{\parindent}{2em} The event horizon of a non-stationary Kinnersley black hole described by the expression~(\ref{eq2}) satisfies the null hypersurface condition
\begin{eqnarray}
\label{eq27}
g^{\mu \nu } \frac{{\partial F}}{{\partial x^\mu  }}\frac{{\partial F}}{{\partial x^\nu  }} = 0.
\end{eqnarray}
Because of the acceleration direction of a non-uniformly rectilinearly accelerating black hole is always pointing to the north pole, namely $\theta  = 0$, the space-time described by Eq.~(\ref{eq2}) is axial symmetry. From the expression~(\ref{eq9}) and Eq.~(\ref{eq27}), we can get the null hypersurface equation that the black hole event horizon satisfying
\begin{eqnarray}
r_H^2  - 2\dot r_H r_H^2  - 2a r_H^3 \cos \theta  - 2M r_H  - 2a r'_H r_H^2 \sin \theta + {r'_H}^2  = 0,
\label{eq28}
\end{eqnarray}
where$\dot r_H  = {{\partial r} \mathord{\left/ {\vphantom {{\partial r} {\partial \upsilon }}} \right. \kern-\nulldelimiterspace} {\partial \upsilon }}
$, $r'_H  = {{\partial r} \mathord{\left/ {\vphantom {{\partial r} {\partial \theta }}} \right. \kern-\nulldelimiterspace} {\partial \theta }}$ . From Eq.~(\ref{eq28}) we can see that, the event horizon $r_H$ is a function which depends not only on time $\upsilon$, but also on angle $\theta$. In order to calculate the temperature of a black hole, here we introduce a new generalized tortoise coordinate transformation \cite{16}
\begin{eqnarray}
r_*  = r + r_H \left( {\upsilon _0 ,\theta _0 } \right) \ln \left[ \frac{ {r - r_H \left( {\upsilon ,\theta } \right)}}{r_H \left( {\upsilon ,\theta } \right)} \right]^{\alpha},\upsilon _*  = \upsilon  - \upsilon _0 ,\theta _*  = \theta  - \theta _0,
\label{eq29}
\end{eqnarray}
in which $\alpha$ is an adjustable coefficient, $\upsilon _0$ and $\theta _0$ are arbitrary constants. The differential forms for above expression become
\begin{eqnarray}
\frac{\partial }{{\partial r}} = \left[ {1 + \frac{\alpha r_H \left( {\upsilon _0 ,\theta _0 } \right)}{{ \left( {r - r_H } \right)}}} \right]\frac{\partial }{{\partial r_* }},
\frac{\partial }{{\partial \upsilon }} = \frac{\partial }{{\partial \upsilon _* }} - \frac{{\alpha r_H \left( {\upsilon_0 ,\theta_0 } \right) r \dot r_H }}{{ r_H \left( {r - r_H } \right)}}\frac{\partial }{{\partial r_* }},
\frac{\partial }{{\partial \theta }} = \frac{\partial }{{\partial \theta _* }} - \frac{ \alpha r_H \left( {\upsilon_0 ,\theta_0 } \right) r {r'_H }}{{ r_H \left( {r - r_H } \right)}}\frac{\partial }{{\partial r_* }}.
\label{eq30}
\end{eqnarray}
Substituting Eq.~(\ref{eq30}) into the Eq.~(\ref{eq21}), we have
\begin{flalign}
& g^{11} \left[ {1 + \frac{\alpha r_H \left( {\upsilon _0 ,\theta _0 } \right)}{{ \left( {r - r_H } \right)}}} \right]^2 \left( {\frac{{\partial S}}{{\partial r_* }}} \right)^2  + 2g^{01} \left[ {1 + \frac{\alpha r_H \left( {\upsilon _0 ,\theta _0 } \right)}{{ \left( {r - r_H } \right)}}} \right] \frac{{\partial S}}{{\partial r_* }}  \left[  \frac{\partial S }{{\partial \upsilon _* }} - \frac{{\alpha r_H \left( {\upsilon_0 ,\theta_0 } \right) r \dot r_H }}{{ r_H \left( {r - r_H } \right)}}\frac{\partial S }{{\partial r_* }} \right]
\nonumber \\
& + 2g^{12} \left[ {1 + \frac{\alpha r_H \left( {\upsilon _0 ,\theta _0 } \right)}{{ \left( {r - r_H } \right)}}} \right] \frac{{\partial S}}{{\partial r_* }}\left[ {P_\theta   - \frac{ \alpha r_H \left( {\upsilon_0 ,\theta_0 } \right) r {r'_H }}{{ r_H \left( {r - r_H } \right)}} \frac{{\partial S}}{{\partial r_* }}} \right] + g^{22} \left[ {P_\theta   - \frac{ \alpha r_H \left( {\upsilon_0 ,\theta_0 } \right) r {r'_H }}{{ r_H \left( {r - r_H } \right)}} \frac{{\partial S}}{{\partial r_* }}} \right]^2
 \nonumber \\
& +g^{33} j^2  + m_0^2  = 0,
\label{eq31}
\end{flalign}
In the above expression, we define $P_\theta   = {{\partial S} \mathord{\left/ {\vphantom {{\partial S} {\partial \theta _* }}} \right. \kern-\nulldelimiterspace} {\partial \theta _* }}$ and $P_\varphi   = {{\partial S} \mathord{\left/ {\vphantom {{\partial S} {\partial \varphi }}} \right. \kern-\nulldelimiterspace} {\partial \varphi }} = j$. $P_\theta$ stands for the generalized momentum, $j$ is a constant that is related to the Killing $\left( {{\partial  \mathord{\left/ {\vphantom {\partial  {\partial \varphi }}} \right. \kern-\nulldelimiterspace} {\partial \varphi }}} \right)$ vector \cite{5}. After processing the above expression, we can get
\begin{flalign}
& \frac{{g^{11} r_H^2 \mathcal{R}^2  + \alpha r r'_H r_H\left( {\upsilon_0 ,\theta_0 } \right) \left\{ {g^{22} \alpha r r'_H r_H\left( {\upsilon_0 ,\theta_0 } \right) - 2g^{12} r_H \mathcal{R}} \right\} - 2 g^{01} \alpha r r_H r_H\left( {\upsilon_0 ,\theta_0 } \right) \mathcal{R} \dot r_H }}{{g^{01} r_H^2 \left( {r - r_H } \right)\mathcal{R} }}\left( {\frac{{\partial S}}{{\partial r_* }}} \right)^2
 \nonumber \\
&+  2\frac{{\partial S}}{{\partial r_* }}\frac{{\partial S}}{{\partial \upsilon _* }} - \frac{{2\left\{ {g^{22} \alpha r r'_H  r_H\left( {\upsilon_0 ,\theta_0 } \right)  - g^{12} r_H \mathcal{R}} \right\}}}{{g^{01} r_H \mathcal{R}}}P_\theta  \frac{{\partial S}}{{\partial r_* }} + \frac{{ {r - r_H } }}{{g^{01} \mathcal{R}}} g^{33} j^2 +  \frac{{{r - r_H } }}{{g_{01} \mathcal{R}}} g^{22} P_\theta ^2  + \frac{{ {r - r_H } }}{{g_{01} \mathcal{R}}}m_0^2  = 0,
\label{eq32}
\end{flalign}
where, $\mathcal{R} =  r - r_H + \alpha r_H\left( {\upsilon_0 ,\theta_0 } \right)$.
According to the conformally flat condition, here we ask that the limit of the coefficients of the $r_H$ in Eq.~(\ref{eq32}) is $1$ when the radius of the black hole approaches the event horizon $r_H$, namely
\begin{eqnarray}
\mathop {\lim }\limits_{\scriptstyle r \to r_H \hfill \atop
  {\scriptstyle \upsilon  \to \upsilon _0 \hfill \atop
  \scriptstyle \theta  \to \theta _0 \hfill}}  \frac{{g^{11} r_H^2 \mathcal{R}^2  + \alpha r r'_H r_H\left( {\upsilon_0 ,\theta_0 } \right) \left\{ {g^{22} \alpha r r'_H r_H\left( {\upsilon_0 ,\theta_0 } \right) - 2g^{12} r_H \mathcal{R}} \right\} - 2 g^{01} \alpha r r_H r_H\left( {\upsilon_0 ,\theta_0 } \right) \mathcal{R} \dot r_H }}{{g^{01} r_H^2 \left( {r - r_H } \right)\mathcal{R} }}
  \nonumber \\
  =  r_H\left( {\upsilon_0 ,\theta_0 } \right).
\label{eq33}
\end{eqnarray}
But we can find that, when $r \to r_H$, an infinite limit of $0/0$-type arises from the above expression. In order to study the tunneling radiation on the event horizon, we use the L'Hospital rule, thus we have
\begin{eqnarray}
\alpha  = \frac{r_H \left[4M + r_H (-2 + 2 \dot r_H + r_H\left( {\upsilon_0 ,\theta_0 } \right) ) + 2 a r_H ( 2 r_H \cos \theta + r'_H \sin \theta ) \right]}{2 r_H\left( {\upsilon_0 ,\theta_0 } \right) \left[ M- r_H \dot r_H - a r_H \left( r_H \cos \theta - r'_H \sin \theta \right)\right]}.
\label{eq34}
\end{eqnarray}
In the above equation $\kappa$ is the coefficient of surface gravity of a variably rectilinearly accelerating black hole. Therefore, on the event horizon, Eq.~(\ref{eq32}) can be expressed as
\begin{eqnarray}
\left( {\frac{{\partial S}}{{\partial r_* }}} \right)^2  + \frac{2}{r_H\left( {\upsilon_0 ,\theta_0 } \right)} \frac{{\partial S}}{{\partial r_* }}\frac{{\partial S}}{{\partial \upsilon _* }} + \frac{AP_\theta}{r_H\left( {\upsilon_0 ,\theta_0 } \right)}  \frac{{\partial S}}{{\partial r_* }} = 0,
\label{eq35}
\end{eqnarray}
where the limit of the coefficient $A$ is
\begin{eqnarray}
\left. A \right|_{r \to r_H }  = \tilde A = \left. {{{2\left( {g_{12}  - g_{22} \alpha r'_H } \right)} \mathord{\left/
 {\vphantom {{2\left( {g_{12}  - g_{22} r'_H } \right)} {g_{01} }}} \right.
 \kern-\nulldelimiterspace} \left({g_{01} \alpha }\right)}} \right|_{\scriptstyle \upsilon  \to \upsilon _0  \hfill \atop
  \scriptstyle \theta  \to \theta _0  \hfill}.
\label{eq36}
\end{eqnarray}
By simplifying Eq.~(\ref{eq35}), we can get
\begin{eqnarray}
\left( {\frac{{\partial S}}{{\partial r_* }}} \right)^2  + 2\frac{{\partial S}}{{\partial r_* }}\left( \frac {\omega  + \omega _0 } {r_H\left( {\upsilon_0 ,\theta_0 } \right)}\right) = 0.
\label{eq37}
\end{eqnarray}
In Eq.~(\ref{eq37}), we order $\omega _0  = {{AP_\theta  } \mathord{\left/ {\vphantom {{AP_\theta  } 2}} \right. \kern-\nulldelimiterspace} 2}$.  At the same time, in the previous work, we proved the relationship ${{\partial S} \mathord{\left/ {\vphantom {{\partial S} {\partial \upsilon _* }}} \right. \kern-\nulldelimiterspace} {\partial \upsilon _* }} = \omega$, where $\omega$ is the energy of the radiation particles. So, by solving the above equation, we can get
\begin{eqnarray}
{{\partial S} \mathord{\left/
 {\vphantom {{\partial S} {\partial r}}} \right.
 \kern-\nulldelimiterspace} {\partial r}} = \left[ {1 + \frac {\alpha r_H\left( {\upsilon_0 ,\theta_0 } \right)}{r-r_H}} \right]{{\partial S} \mathord{\left/
 {\vphantom {{\partial S} {\partial r_* }}} \right.
 \kern-\nulldelimiterspace} {\partial r_* }} = \left[ {1 + \frac {\alpha r_H\left( {\upsilon_0 ,\theta_0 } \right)}{r-r_H}} \right] {\left[ \frac{{\left( {\omega  - \omega _0 } \right) \pm \left( {\omega  - \omega _0 } \right)}}{r_H\left( {\upsilon_0 ,\theta_0 } \right)} \right]}.
\label{eq38}
\end{eqnarray}
In the expression, it can be found that there is only one singularity, that is $r_H$. After integration, the result can be expressed as
\begin{eqnarray}
S_ \pm   = \int {\left[ {1 + \frac {\alpha r_H\left( {\upsilon_0 ,\theta_0 } \right)}{r-r_H}} \right] {\left[ \frac{{\left( {\omega  - \omega _0 } \right) \pm \left( {\omega  - \omega _0 } \right)}}{r_H\left( {\upsilon_0 ,\theta_0 } \right)} \right]} dr}  = {{i\pi \alpha \left[ {\left( {\omega  - \omega _0 } \right) \pm \left( {\omega  - \omega _0 } \right)} \right]}  },
\label{eq39}
\end{eqnarray}
where $ + \left(  -  \right)$ is a outgoing (ingoing) solution. In order to obtain the tunneling rate, we have to consider both the outgoing and the ingoing solutions. As a result, the tunneling rate of the particle becomes
\begin{eqnarray}
\Gamma  = \frac{{\Gamma _{out} }}{{\Gamma _{in} }} = \frac{{\exp \left( { - 2{\mathop{\rm Im}\nolimits} S_ +  } \right)}}{{\exp \left( { - 2{\mathop{\rm Im}\nolimits} S_ -  } \right)}} = \exp \left[ {{{- 4 \pi \alpha \left( {\omega  - \omega _0 } \right)}}} \right].
\label{eq40}
\end{eqnarray}
Using the Boltzmann factor expression, the Hawking temperature of the black hole is given by
\begin{eqnarray}
T_H  = \frac{1 }{{4 \pi \alpha}} = \frac {1}{4 \pi}\frac{{2 r_H\left( {\upsilon_0 ,\theta_0 } \right) \left[ M- r_H \dot r_H - a r_H \left( r_H \cos \theta - r'_H \sin \theta \right)\right]}}{{r_H \left[4M + r_H (-2 + 2 \dot r_H + r_H\left( {\upsilon_0 ,\theta_0 } \right) ) + 2 a r_H ( 2 r_H \cos \theta + r'_H \sin \theta ) \right]}}.
\label{eq41}
\end{eqnarray}
In the above expression, while ignoring the angular momentum $a$ and the coefficients $r'_H$, $\dot r_H$, meanwhile setting $r_H=2M$, we can find that the surface gravity of a dynamic Kinnersley black hole reduces to the case of Schwarzschild hole. Therefore, Eq.~(\ref{eq41}) also reduces to the Hawking temperature of a Schwarzschild black hole. Furthermore, it is well-known that the relationship between the temperature and the surface gravity is $T=\kappa /(2 \pi)$, where the $\kappa$ is the surface gravity of the black hole. Combined with this fact, we find from Eq.(\ref{eq41}) that the value of $\alpha$ is equal to $ 1 / ( 2 \kappa)$.

In the calculation of the above process, by using the Hamilton-Jacobi equation, we study the tunneling behavior of spin $1/2$ and $3/2$ particles in a dynamic Kinnersley black hole, and get the Hawking temperature of the black hole. In fact, in previous work, we have proved that the Hamilton-Jacobi equation can describe the movement behavior in curved spacetime for scalar particles with spin $0$, fermions with spin $1/2$ and $3/2$, boson with spin $1$ and graviton \cite{13}. For instance, the Klein-Gordon equation for the scalar particle can be described by
\begin{eqnarray}
\frac{1}{\sqrt{-g}} \frac{\partial}{\partial x^{\mu}} \left(\sqrt{-g} g^{\mu \nu} \frac{\partial}{\partial x^{\mu}} \right) \Psi - \frac{m^2}{\hbar^2} \Psi=0.
\label{eq42}
\end{eqnarray}
In a similar way, after substituting the wave function
\begin{eqnarray}
\Psi  = \zeta \exp \left( {\frac{i}{\hbar }S} \right),
\label{eq43}
\end{eqnarray}
into Eq.(\ref{eq42}), and neglecting the terms with $\hbar$, then it is easy for us to obtain the Hamilton-Jacobi equation
\begin{eqnarray}
g^{\mu \nu } \left( {{{\partial S} \mathord{\left/ {\vphantom {{\partial S} {\partial \mu }}} \right. \kern-\nulldelimiterspace} {\partial \mu }}} \right)\left( {{{\partial S} \mathord{\left/
 {\vphantom {{\partial S} {\partial \nu }}} \right. \kern-\nulldelimiterspace} {\partial \nu }}} \right) + m^2  = 0.
\label{eq44}
\end{eqnarray}
This means, by using the Hamilton-Jacobi equation, the tunneling rate and the Hawking temperature for scalar particles can also be addressed in the non-stationary black hole. In a word, we confirm from Eq.(\ref{eq21},\ref{eq26},\ref{eq42},\ref{eq44}) that the Hamilton-Jacobi equation is a basic equation, and it can be used to study the tunneling behavior of any particle in curved space-time.

\section{Conclusions}
 \label{section4}
\setlength{\parindent}{2em} In this paper, starting from both Dirac equation of spin $1/2$ fermion and Rarita-Schwinger equation of spin $3/2$ fermion, we can obtain a Hamilton-Jacobi equation. Moreover, making use of this equation, we investigate the fermion tunneling rate and the Hawking temperature of a non-stationary Kinnersly black hole. According to our research conclusion, the Hamilton-Jacobi equation can be derived from the kinetic equations of arbitrary spin particles. This shows that the Hamilton-Jacobi equation is a basic semi-classical equation, which can be used to study the quantum tunneling behavior of arbitrary spin particles. Furthermore, through the comparison of the previous work, we can obviously see that, it is very convenient to study tunneling radiation by using the Hamilton-Jacobi equation. Especially for the fermion tunneling, since there is no need to construct complex gamma matrix for tedious calculation, which greatly reduces the workload. It is better for people to carry out the creative research in depth.

\vspace*{3.0ex}
{\bf Acknowledgements}
\vspace*{1.0ex}
This work is supported by the Natural Science Foundation of China (Grant No. 11573022).

\end{document}